\documentclass{jaa}
\usepackage{natbib}
\usepackage{graphicx}
\usepackage{siunitx}
\usepackage{url}
\usepackage{xcolor, soul}
\definecolor{columbiablue}{rgb}{0.61, 0.87, 1.0}
\sethlcolor{columbiablue}

\begin{document}\sloppy

%%paper title
%%For line breaks \\ can be used within title
\title{Level correlations of the CMB temperature angular power spectrum}

%%author names are separated by comma (,)
%%use \and before the last author name
%%use a * along with the number separated by comma
%% for the  author for correspondence
%%\textsuperscript{number} is used for affiliation
%%\affilOne, \affilTwo etc., upto \affilTwentyfive is possible
%%Please note the first letter after \affil is capitalised in the command
%%

\author{Md Ishaque Khan\textsuperscript{1}, Rajib Saha\textsuperscript{1*}}
\affilOne{\textsuperscript{1}Department of Physics, Indian Institute of Science Education and Research, Bhopal-462066, India\\}
%\affilTwo{\textsuperscript{2}Department of Q, University Z, Place Pincode, Country.}

%%escape two column mode for title, affiliation and abstract
%%by giving \twocolumn command as shown

\twocolumn[{

\maketitle

%%include \corres to print the corresponding author Email id
\corres{rajib@iiserb.ac.in}

%%include \msinfo for
%%manuscript information such as
%%received, revised and accepted dates
%%
\msinfo{}{}

%%abstract
\begin{abstract}
If the principle of statistical isotropy is valid, then the angular power spectrum (APS) of cosmic microwave background (CMB) radiation is uncorrelated between different multipoles. We propose a novel technique to analyse any possible correlations of the foreground cleaned CMB temperature angular power spectrum (APS) measures ($C_{\ell}$ and $\mathcal{D}_\ell=\frac{\ell(\ell+1)}{2\pi}C_\ell$). This is motivated by the behaviour of level spacings between random matrix eigenvalues. The method helps distinguish uncorrelated statistically isotropic CMB APS from correlated APS, where the latter can arise due to breakdown of isotropy or presence of some residual systematics in the foreground cleaned CMB maps. Spacings of statistically isotropic CMB $C_\ell$'s and $\mathcal{D}_\ell$'s are seen to closely obey Poisson statistics and introduction of correlations changes the distribution to appropriate Wigner-Dyson statistics. For foreground cleaned CMB, we employ the average spacing of consecutive multipole APS for multipoles $\in [2,31]$. This estimator is sensitive to departures from the null hypothesis of zero correlations between the APS measures of statistically isotropic CMB. We study full sky WMAP 9 year ILC and 2018 Planck foreground cleaned maps (Commander, NILC and SMICA). Sans parity distinctions, average spacings are in good agreement with theoretical expectation. With parity distinctions, even multipoles indicate unusually low average spacings for both $C_\ell$'s (at $\geq 98.86\%$ C.L.) and $\mathcal{D}_\ell$'s (at $\geq 95.07\%$ C.L.). We use an inpainting method based on constrained Gaussian realisations and show that for the Planck $U73$ and WMAP $KQ75$ masks, all the foreground cleaned inpainted CMB maps robustly confirm the existence of such unusually low average spacings of even multipole APS. In addition, this signal is independent of the non-Gaussian cold spot.
\end{abstract}

%%insert keywords separated by 3 hyphens using \keywords{words}
\keywords{Cosmology---CMB Anisotropy---Statistical Anisotropy---Wigner Surmise}

}]
%%close the twocolumn escape here

%%include \doinum{number}for the DOI number in the header
%%include \volnum{number} for the volume number in the header
%%include \year{yyyy} for  year of publication in the header
%%include \pgrange{num--num} page range of article in the header
%%include \artcitid{num} for the article citation id
%%include \lp to print last page of the article
%%include \setcounter{page}{pagenum} for the exact starting page of the article

\doinum{}
\artcitid{}
\volnum{000}
\year{0000}
\pgrange{1--16}
\setcounter{page}{1}
\lp{16}

\section{Introduction}\label{sec:intro}

Large scale fluctuations in the CMB temperature potentially encode information about inflation \mbox{\citep{CHENG2014140}}, and hence any signatures that might have arisen primordially. These fluctuations are expected to be statistically isotropic since the power spectrum of the quantum perturbations is hypothesised to be rotationally invariant. This directly implies that the angular power spectrum (APS) measures are uncorrelated between different multipoles. A violation of the assumption of statistically isotropic CMB temperature field then manifests in the form of correlations in the APS measures \citep{Chang_2018, PhysRevD.91.023515}.

In this work, we propose a novel method, that has not been explored in existing literature, to study correlations among the temperature fluctuations of the CMB. The method investigates the spacings between the CMB APS measures, namely $C_\ell$'s and $\mathcal{D}_\ell$'s ($=\frac{\ell(\ell+1)}{2\pi}C_\ell$'s) \footnote{Here, apostrophes (') are used to denote plural forms.}. The principal objective of the current article is to detect any signature of correlations in the angular power spectra of the foreground minimized CMB maps following the methodology of hypothesis testing. The null hypothesis corresponds to the assertion that the CMB angular power spectrum from the foreground cleaned CMB maps are uncorrelated. This is an important scientific question to ask since if the null hypothesis can be invalidated it may warrant new physics if the correlations are generated due to any small level of a primordial signal. If the correlations are induced by the residual foregrounds or any other systematic effects present in the cleaned maps, then special care must be adopted in using these cleaned maps in cosmological parameter estimation for accurate interpretations of these variables.  Some unknown systematics may creep in during the analysis pipeline of satellite data collection and/or during the map making process.  These discussions illustrate the fundamental importance of the research work carried out in this work.

To motivate more into the focus of the current research work in the context of studies of CMB anisotropies   we note that, in existing literature, the deficit of large angle correlation was seen for COBE-DMR four year maps \citep{Hinshaw_1996} and subsequently for the WMAP first year data \citep{Bennett_2003_2}. Thereafter, almost negligible correlation was seen above $\ang{60}$ for later WMAP and Planck releases  \citep{10.1111/j.1365-2966.2009.15270.x, PhysRevD.75.023507, 10.1093/mnras/stv1143}, with increasing statistical significance. Hence it was argued to be a truly anomalous feature instead of being causative of a specific a posteriori choice of statistic \citep{10.1111/j.1365-2966.2009.15270.x}. In addition, this anomaly has recently been shown to exist in the latest Planck polarisation data \citep{Chiocchetta_2021}. Besides, parity asymmetry in the APS was found \citep{Kim_2010, PhysRevD.82.063002} and it was shown that the anomaly disappears without the contribution from first six low multipoles ($\ell=2,...,7$) \citep{10.1111/j.1365-2966.2011.19981.x}. Later \cite{kim_lack_2011} also showed that the parity asymmetry in the APS is phenomenologically equivalent to deficit of large angle correlation.

Thus, in literature, there are independent studies of (a) the deficit of large angle correlation and (b) its equivalence with odd-parity preference of the APS. However, there exist no investigations in the literature regarding whether there are any unusual correlations in either even or odd or both parities of multipoles of APS when their APS estimators ($C_\ell$'s and related $\mathcal{D}_\ell$'s) are separately considered instead of the two point angular correlation function. Further, the authors of \cite{PhysRevD.75.023507} have noted that any covariance between $C_\ell$'s must be studied in addition to the two point angular correlation function. Besides, understanding correlations among the APS is important as these are directly used for cosmological parameter estimation. The nature of correlations among separate sets of odd multipole and even multipole APS as well as the complete set, can be captured by their respective average spacing estimators (Section \ref{estim}) that we have devised as a novel statistic for our study.

Several studies can be found in existing literature that claim signatures of statistical anisotropy in the foreground cleaned CMB maps \citep{PhysRevD.69.063516, doi:10.1142/S0218271804005948, 10.1111/j.1365-2966.2009.14728.x, PhysRevLett.93.221301, 10.1111/j.1365-2966.2008.12960.x, PhysRevLett.95.071301, PhysRevD.89.023010}. Besides, theoretically anisotropic models over large scales have been proposed to account for some of such observed statistically anisotropic signatures \citep{PhysRevD.75.083502, MALEKNEJAD2013161,soda_statistical_2012, 10.1093/mnras/sty1689, li_finslerian_2017, dimastrogiovanni_non-gaussianity_2010}. In the past, analysis of CMB data unveiled many anomalies in the foreground minimized maps of WMAP \citep{Copi:2010na} and Planck \citep{Schwarz_2016} satellite missions. This has entailed the study of anomalies in both WMAP and Planck maps to make departures more discernible if any such exist. Such anomalies include the north-south power asymmetry \citep{Bernui_2014}, which is relatively insignificant for all scales from Planck \citep{Quartin_2015}, high degree of octupole-quadrupole alignment \citep{Tegmark:2003ve,deOliveira-Costa:2003utu,Schwarz:2004gk} that strengthens on removing the frequency dependent kinetic Doppler quadrupole \citep{Notari_2015}, quadrupole power deficit \citep{Bennett_2003,10.1046/j.1365-2966.2003.07067.x} and planarity of the octupole \citep{PhysRevD.69.063516}, the power excess for lower odd multipoles \citep{PhysRevD.72.101302}, the non-Gaussian cold spot \citep{Vielva_2004,10.1111/j.1365-2966.2004.08419.x,Cruz_2007}, unusually weak non-uniformity in the placement of CMB hot and cold spots \citep{Khan_2022}, and the like.

Foreground cleaned CMB maps are obtained upon application of elaborate cleaning methods \citep{2014A&A...571A..12P, 10.1111/j.1365-2966.2011.20182.x, cardoso:in2p3-00022125, 10.1111/j.1365-2966.2003.07069.x, 2005EJASP2005.2400B, Tegmark:1995pn, 1994ApJ...432L..75B, 2011_7_yr_wmap, Saha_2011, Sudevan_2020, Yadav_2021, 10.1093/mnras/staa3935} such as those of Gibbs sampling \citep{2008ApJ...676...10E}, Spectral Matching Independent Component Analysis \citep{2007astro.ph..2198D}, Internal linear combination (ILC) in needlet space \citep{10.1111/j.1365-2966.2011.19770.x}, and ILC in pixel space \citep{2003ApJS..148...97B} to the foreground contaminated maps of CMB radiation observed at different frequencies. In the case of foreground cleaned CMB maps, a breakdown of statistical isotropy may be caused due to primordial features which are unaccounted for in the concordance ($\Lambda CDM$) model of cosmology, or due to unaccounted agents between the surface of last scattering and the observer, or any possible residual foregrounds left over after cleaning, or due to some minor systematics that may have crept in during the analysis pipeline of satellite data collection and/or the map making procedure.

Our study here focuses on an effort to discover unusual signatures of correlations between consecutive multipole APS as manifested by their spacings. Our paper is organised as follows. In Section \ref{bvar}, we discuss about the statistics of Poisson and Wigner-Dyson in context of level spacings, and describe the basic variables, i.e, the APS spacings, used for our study. In Section \ref{trans} we elucidate level clustering and repulsion for CMB APS. Section \ref{estim} specifies the average spacing estimator as the novel statistic devised for our study of foreground cleaned CMB maps. Section \ref{metho} describes our way of testing consecutive multipole spacings with simulations. We report our primary results in Section \ref{results} for all multipole and even/odd multipole spacings using the average spacing estimator. In Section \ref{inpainted} we establish the robustness of our findings with inpainted maps. And, in Section \ref{conclusion} we summarise these results and state our inferences.

\section{Statistics and Basic Variables}\label{bvar}

\begin{figure*}%[18]{r}{0.6\textwidth}%,\columnwidth}%[h!]
    \centering
        
    \includegraphics[width=1.5\columnwidth]{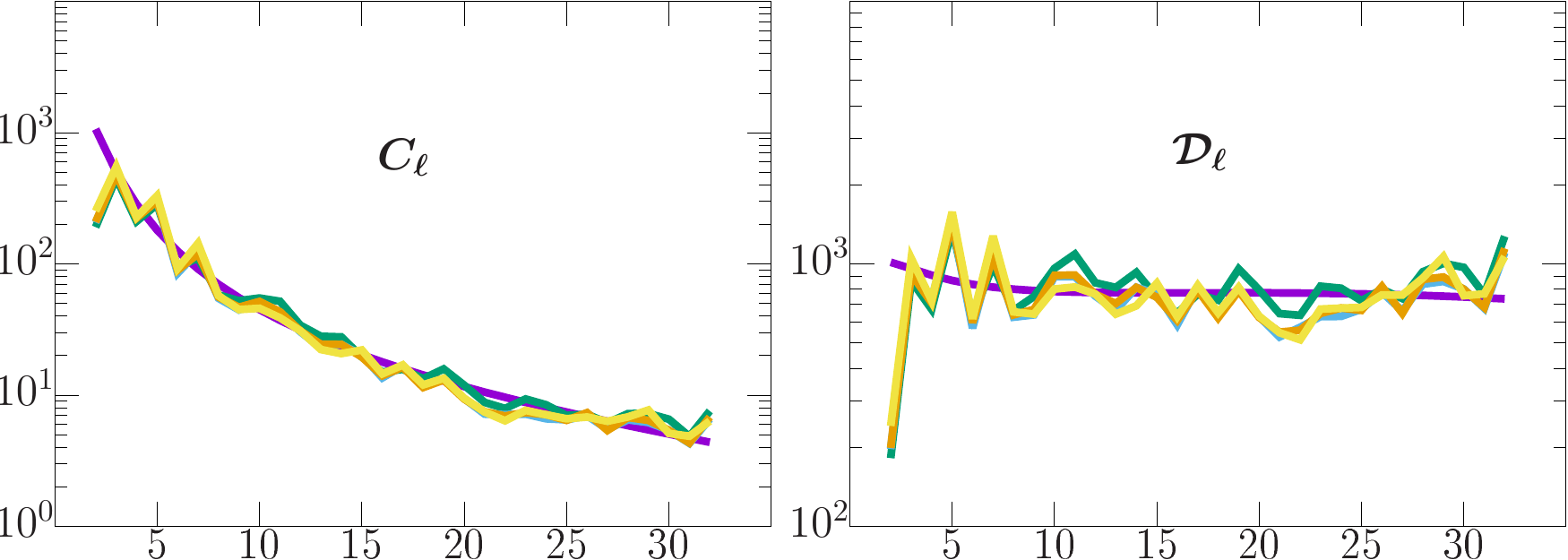}
    \caption{$C_\ell$ and $\mathcal{D}_\ell$ versus $\ell$: the subfigures compare the theoretical APS best fit to Planck 2018 data (purple) with estimated APS from COMM (green), NILC (light blue), SMICA (ochre) and WMAP (yellow) maps used here. The monopole and dipole have been excluded in this figure. The vertical axis is in log-scale.}
    \label{fig_ll1cl}
    
%  \begin{spacing}{1.0}
  
  %\end{spacing}
\end{figure*}

In this section we review the relevant statistical distributions and the basic variables of our work. The analogue of `level spacings' in the form of absolute differences of the consecutive multipole APS measures are given by,
\begin{eqnarray}\label{eq_sp}
   \Delta C_\ell&=&|C_{\ell_1}-C_{\ell_2}|, \nonumber \\
   \Delta \mathcal{D}_\ell&=&\left|\frac{\ell_1(\ell_1+1)}{2\pi}C_{\ell_1}-\frac{\ell_2(\ell_2+1)}{2\pi}C_{\ell_2}\right|,
\end{eqnarray} where, $\ell_1,\ell_2$ are consecutive multipoles. Assuming statistical isotropy, $C_\ell$'s represent the unbiased APS estimators defined as \citep{Hinshaw_2003},
\begin{eqnarray}\label{cl_alm}
    {C_\ell}&=&\frac{1}{2\ell+1}\sum_{m=-\ell}^{\ell} |a_{\ell m}|^2,
   \end{eqnarray}
and, the $a_{\ell m}$'s are coefficients of the expansion,
\begin{eqnarray}
    \Delta T(\hat n)&=&\sum_{\ell}\sum_{m=-\ell}^\ell a_{\ell m}Y_{\ell m}(\hat n).
\end{eqnarray}$\Delta T(\hat n)$'s are the temperature anisotropies relative to the uniform mean temperature of nearly $2.726K$ \citep{Fixsen_2009} of the CMB. Here $\hat n$ denotes a direction in the sky. The APS measures used in this work are of low multipoles ($\ell\in[2,31]$) corresponding to large angular scales. For reference, we show the two natural APS measures, $C_\ell$'s and $\mathcal{D}_\ell$'s in Figure \ref{fig_ll1cl}. As $\mathcal{D}_\ell$ fluctuates about a nearly constant curve over large scales, it is interesting to study spacings of the same in addition to $C_\ell$'s.

Since $C_\ell$'s arise from independent and uncorrelated Gaussian random $a_{\ell m}$'s \eqref{cl_alm}, they should be uncorrelated. As $\mathcal{D}_\ell$'s differ from $C_\ell$'s only by a multiplicative factor of $\frac{\ell(\ell+1)}{2\pi}$, they are expected to be uncorrelated as well. Therefore, low spacings (level clustering) for the $C_\ell$'s and $\mathcal{D}_\ell$'s may be favoured because of Poisson statistics \citep{proceedings_1977}, which holds for i.i.d. variables. However, there will be some deviations from purely Poisson nature for spacings of statistically isotropic realisations of the CMB having uncorrelated APS, since they are constrained by cosmic variance, and are not identically distributed \footnote{ Each $C_\ell$ is a $\chi^2$ variable but with $2\ell+1$ degrees of freedom. Thus the closest to nearly identical distributions for any two $C_\ell$'s can be considered by taking spacings of consecutive multipoles, as done here. This minimizes the difference between the distributions of the two $C_\ell$'s in a spacing to that of $(2(\ell+1)+1)-(2\ell+1)=2$ degrees of freedom.}. If the observed $C_\ell$'s were specially correlated, the Wigner-Dyson statistics \citep{PhysRevLett.52.1}  which favour level repulsion could effectively describe these spacings, and preclude the possibility of consecutive $C_\ell$'s and $\mathcal{D}_\ell$'s being arbitrarily close.

It is known that Poisson statistics (describing a possibly integrable underlying system) and Wigner-Dyson statistics  (describing those with classically chaotic or non-integrable counterparts) can be represented as in equation \eqref{eqns} for a spacing $s$ between two consecutive eigenvalues of the associated random matrix, with a probability distribution $p(s)$ \citep{Meh2004}.
\begin{eqnarray}\label{eqns}
    p(s)&\propto& e^{-s} \quad \text{(Poisson statistics)}, \nonumber \\
   p(s)&\propto& s^\alpha e^{-b_\alpha s^2} \quad \text{(Wigner-Dyson statistics)},
\end{eqnarray} (where, $\alpha>0$). There can be a large class of random matrices, which may have i.i.d entries extracted from distributions other than those of the standard Gaussian ensembles. It has been seen for such matrices \citep{PhysRevResearch.2.023286, 10.1214/15-AAP1129} as well that their eigenvalue spacings obey some appropriate Wigner-Dyson form in the presence of correlations. This is the concept of universality \citep{GUHR1998189}, which we seek to explore in the context of the APS measures.

Due to the parity inversion property of spherical harmonics \footnote{Under a parity transform, i.e., $ \hat n \to - \hat n, Y_{\ell m}(\hat n) \to Y_{\ell m}(-\hat n)=(-1)^\ell Y_{\ell m}(\hat n) \implies a_{\ell m} \to (-1)^\ell a_{\ell m}$}, the temperature anisotropy field can be reconstructed as a sum of a symmetric and an anti-symmetric function \citep{kim_jaiseung_naselsky_pavel_hansen_martin_2012}. These functions are of even and odd parity, respectively and the power for a multipole range can be rewritten as a sum of contributions from the even and odd multipole APS. Hence, we study nearest neighbour spacings of separate sets of even and odd multipole APS measures in addition to an analysis of APS spacings without a parity distinction. This may help us find if any parity preference of spacings exist, which can indicate deviations from the assumed isotropy of the universe on large scales.

\section{Level clustering and level repulsion in CMB APS}\label{trans}

In this section, we discuss how the transition between level clustering (Poisson statistics) and level repulsion (Wigner-Dyson statistics) for the CMB angular power spectrum (APS) may occur, due to introduction of correlations in otherwise uncorrelated CMB APS of statistically isotropic CMB maps. We demonstrate how the respective phenomena of level clustering and level repulsion are also applicable in the context of CMB APS. Level clustering occurs when the APS is uncorrelated between multipoles for statistically isotropic CMB. Whereas level repulsion takes place when the APS gets correlated on introduction of statistical anisotropy.

To illustrate clustering and repulsion between `levels' of the CMB APS at different multipoles, we utilised some typical foregrounds to introduce correlations in otherwise statistically isotropic CMB maps. However, we must note that there can be several mechanisms by which correlations may be introduced in the CMB APS, such as a small statistically anisotropic primordial signal, any minor residual foregrounds, or other systematics left over due to the analysis pipeline followed in satellite data collection and/or the algorithms used for preparing foreground cleaned CMB maps. Thus, introduction of statistical anisotropy in the CMB APS by addition of foreground contamination is only one of various possible mechanisms, and has been chosen solely as a representative example for demonstrating level clustering in correlated CMB APS. This helps benchmark our methodology and also provides important insights regarding the classification of possible nature of correlations that are detected on the foreground cleaned CMB maps.

Statistically isotropic (SI) CMB maps can be generated as Gaussian random realisations based on the concordance ($\Lambda CDM$) model. Foregrounds introduce statistical anisotropy in the SI CMB realisations. Thus, an addition of foregrounds to SI CMB maps is expected to enhance correlations between $C_\ell$'s that may cause the level spacing distributions to obey an appropriate deviation from level clustering. To quantify such a departure, the gap ratio introduced by \citep{PhysRevB.75.155111}, is:
\begin{eqnarray}
    r_\ell&=&\frac{\min(\Delta_\ell,\Delta_{\ell-1})}{\max(\Delta_\ell,\Delta_{\ell-1})}.
\end{eqnarray}
Here, $\Delta_\ell$ can be either of $\Delta C_\ell$ or $\Delta D_\ell$. The ensemble average of this ratio has standard values for various distributions. For Poisson, $\langle r\rangle\simeq 0.38$, for GOE (Gaussian Orthogonal Ensemble), $\langle r \rangle\simeq 0.5295$. Further, \citep{PhysRevLett.110.084101} calculated the same for GUE (Gaussian Unitary Ensemble) and GSE (Gaussian Symplectic Ensemble), as $\langle r \rangle\simeq 0.60266$, $0.67617$, respectively. The advantage of using the mean gap ratio is that we need not concern ourselves with unfolding, to remove effects of local densities of the variables being investigated for their spacing distribution. Besides, the mean gap ratios may help us assign the appropriate distributions to these spacings, and give a measure of the how chaotic the underlying system is.

As our APS measures $C_\ell$'s and $D_\ell$'s are not i.i.d.,  we may expect slightly deviated forms from the Poisson and Wigner-Dyson behaviours given by their gap ratios of consecutive spacings. We have generated $10^4$ realisations of SI CMB maps based on the theoretical APS best fitted to Planck 2018 data \citep{pl2018}. To these, three typical foregrounds (synchrotron, thermal dust and free-free emission maps obtained by the Commander cleaning algorithm of the Planck Legacy archive \citep{Planck_maps}) were added after being extrapolated at $\SI{100}{GHz}$, to obtain $10^4$ statistically anisotropic (SA) CMB maps.

\begin{figure*}
    \centering
    \includegraphics[width=1.5\columnwidth]{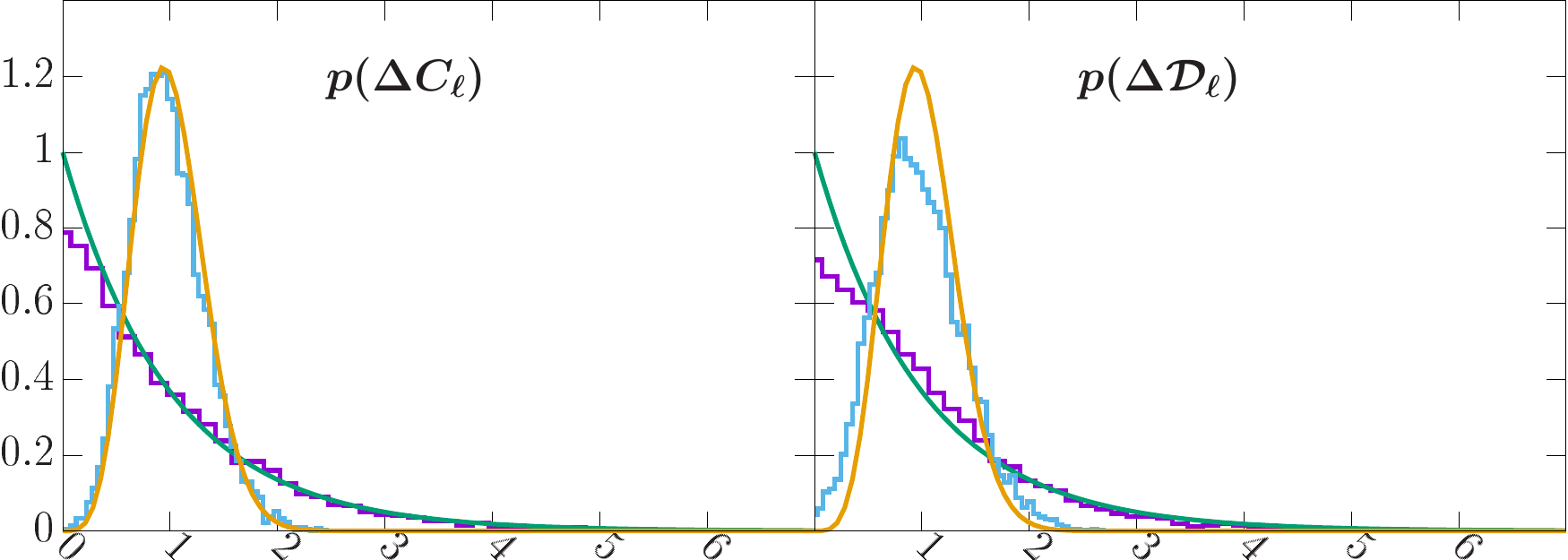}
    \caption{Left panel: Distribution of the $\Delta C_\ell$ spacing from SI CMB maps without any foreground addition (in purple) and with addition of three foregrounds (in light-blue), i.e, synchrotron, thermal dust, and free-free emission extrapolated at $\SI{100}{GHz}$. There is a clear departure from nearly Poisson ($p(s)=\mathrm{e}^{-s}$, green curve) to Wigner-Dyson form of GSE ($p(s) = \frac{2^{18}}{3^6\pi^3}s^4 \mathrm{e}^{-\frac{64}{9\pi} s^2}$, orange curve) for SA CMB maps. Right panel: Distribution of the $\Delta \mathcal{D}_\ell$ spacing from SI (purple) and SA (light-blue) CMB maps. Again a departure from approximately Poisson to some appropriate level repulsion statistics is seen.}
    \label{fig_sp8}
\end{figure*}

From these realisations, we find that $\langle r\rangle=0.41702294$, $0.67770755$, respectively for SI and SA CMB $C_\ell$'s. For spacings of $C_\ell$'s, we can say that roughly Poisson statistics is followed by SI CMB $\Delta C_\ell$'s, whereas for SA CMB, the GSE form of Wigner-Dyson statistics may be appropriate. It is well known that the GSE form is obeyed by underlying systems that do not have rotational symmetry, and the value of the mean gap ratio found for $\Delta C_\ell$'s may be related with the lack of rotational symmetry for SA CMB maps. Likewise, for $\Delta\mathcal{D}_\ell$'s we find $\langle r\rangle=0.45045788$, $0.82103559$ for SI and SA CMB, respectively. These higher values of $\langle r \rangle$ are indicative of additional dependence among $\mathcal{D}_\ell$'s as $\frac{\ell(\ell+1)}{2\pi}\langle C_\ell \rangle\simeq constant$ for low $\ell$'s. In the first panel of Figure \ref{fig_sp8} we have plotted the probability distribution of $\Delta C_\ell$ to illustrate the change in the spacing distribution from approximately Poisson to GSE form of Wigner-Dyson statistics. In the second panel of Figure \ref{fig_sp8}, we see that the curves for $\Delta \mathcal{D}_\ell$'s are not easily classifiable into the standard forms given by Poisson or Wigner-Dyson statistics. However, the subfigure shows the transition between level clustering to level repulsion. This is in agreement with the values of mean gap ratios for $\Delta \mathcal{D}_\ell$'s which show a departure from very low to high correlations, for SI and SA CMB maps.

\section{Level correlation estimator}\label{estim}

We propose a novel estimator for probing correlations among the APS of the foreground cleaned CMB temperature maps. For the range of multipoles $[2,31]$, the spacings between consecutive $C_\ell$'s and $\mathcal{D}_\ell$'s are estimated. Say, such consecutive multipole spacings are  $[\Delta_1,\Delta_2,...]$, then the average spacing estimator is
\begin{eqnarray}\label{estimators}
    avg_i&=&\frac{\Delta_1^i+\Delta_2^i+...}{N}.
\end{eqnarray} Here, $i=a,o,e$, where $a,o,e$ stand for all, odd and even multipoles, and $N=$ number of spacings for the given range of $\ell$'s.

\begin{figure*}%[0]{I}{\columnwidth}%[h!]
 \centering
 \includegraphics[width=1.5\columnwidth]{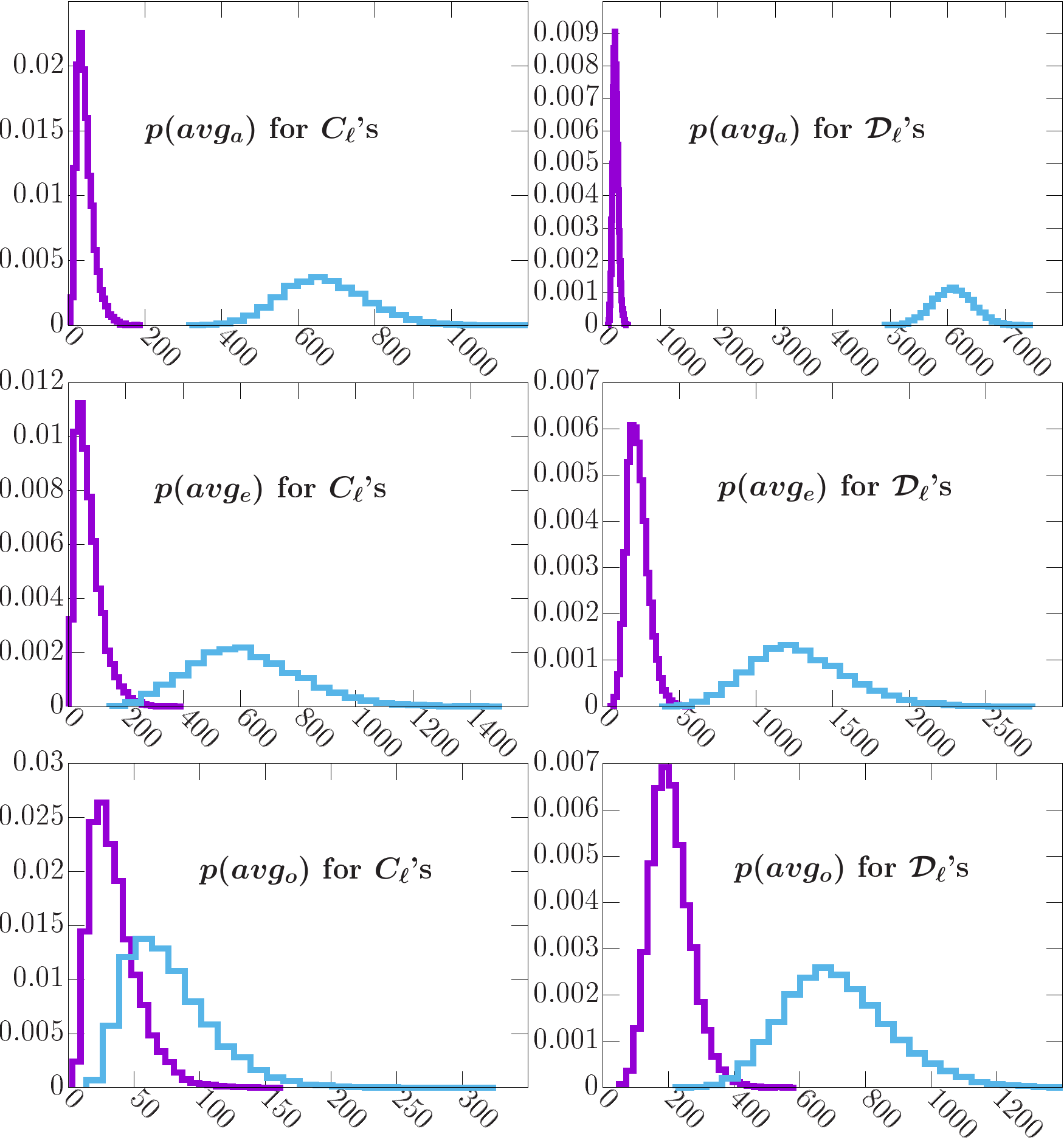}
\caption{Probability distributions of $avg_a$, $avg_e$, and $avg_o$ from $10^4$ statistically isotropic (purple) and anisotropic (light-blue) CMB maps for $C_\ell$'s and $\mathcal{D}_\ell$'s. Foreground impurities added to statistically isotropic CMB maps enhance correlations between consecutive multipole APS and cause average spacings to be larger.}
\label{avg_fg}
\end{figure*}

This estimator helps characterise behaviours of APS spacings and hence large angle correlations in an easily quantifiable way. Besides, the choice of this estimator is a priori. The behaviour of all, even and odd multipole spacings for simulated SI and SA CMB maps, is shown in Figure \ref{avg_fg}. These have been obtained from $10^4$ realisations of SI CMB to which foregrounds were added to obtain $10^4$ SA CMB maps, as described in Section \ref{trans}. The overall effect of addition of foregrounds to statistically isotropic CMB realisations is to shift the average estimator to higher values. Thus it may be highly unlikely to attribute unusually low mean spacings in foreground cleaned CMB maps to any residual foregrounds. The advantage of using this estimator is that we can consider the mean spacing from many statistically isotropic CMB realisations to compare with that from foreground cleaned CMB data. Otherwise, with entities like the gap ratio, any individual spacing, or the Pearson's correlation coefficient, we need an ensemble of realisations, while we have only one universe to observe.

\section{Methodology}\label{metho}

We have compared values of the average spacing estimator for foreground cleaned CMB APS with those of the theoretical APS (Planck 2018 best fit \citep{pl2018}). With the help of the HEALPix \citep{2005ApJ...622..759G} package, we generate $10^4$ statistically isotropic CMB maps of the theoretical temperature APS to account for statistical fluctuations. 

CMB maps used here for probing APS spacings are full-sky foreground cleaned maps of WMAP 9 year ILC and 2018 release full mission Planck Commander (COMM), NILC and SMICA maps from latest sources, i.e., \cite{WMAP_ilc} and \cite{Planck_maps}, respectively. These have been downgraded with the help of HEALPix \citep{2005ApJ...622..759G} software facilities to a HEALPix $n_{side}=16, n\ell_{max}=32$ (hence, an appropriate pixel-window) with no beam smoothing (fwhm{\_}arcmin$=0.0$) and the statistically isotropic CMB maps are obtained using the same resolution. With all foreground cleaned and simulated statistically isotropic CMB maps thereof on an equal footing, we have proceeded with the analysis. We have excluded multipoles $\ell=0,1$ as these correspond respectively to the monopole of uniform CMB temperature ($\approx2.726K$) \citep{Fixsen_2009}, and the dipole which arises due to our peculiar motion relative to the CMB rest frame \citep{Bucher:2015eia}. In addition, we have ignored contributions from noise, as it is not expected to be significant at the large scales studied here \citep{Tegmark_2000}.
 
We have taken the average ($avg_i$) spacing (\ref{estimators}) of $C_\ell$'s and $\mathcal{D}_\ell$'s for consecutive multipoles firstly without any parity distinction and later separately for odd and even multipoles. For example, for the $6$ multipoles in the range $[2,7]$ for $C_\ell$'s, we have $5$ spacings with no parity based distinction namely, $|C_2-C_3|,|C_3-C_4|,...,|C_6-C_7|$. Whereas for even and odd multipoles taken distinctly, in the same range, we have $2$ spacings each for even and odd multipoles i.e., $|C_2-C_4|,|C_4-C_6|$ and $|C_3-C_5|,|C_5-C_7|$ respectively. Thus $avg_a$ is the mean value of $5$ spacings, while $avg_e$, $avg_o$ are mean values of $2$ spacings each. Average spacings for consecutive $\mathcal{D}_\ell$'s are found in a similar fashion. The range of multipoles used for our study is $\ell\in[2,31]$. With this chosen range of multipoles (i.e, $\ell\in[2,31]$), we are able to consider an equal number of odd and even multipole spacings. 
 
For characterising the extent to which the average spacing from foreground cleaned CMB maps may be different from statistically isotropic realisations of the CMB in a quantitative way, we define the probability $P^t(avg_i)$. Here, $i=a,o,e$ which stand for all (no parity based distinction), odd and even multipoles respectively. The fraction $P^t(avg_i)$ is calculated by counting the number of statistically isotropic CMB simulations having the value of the $avg_i$ estimator greater than the foreground cleaned CMB map and dividing the number by the total number of simulations, that being $10^4$ in this paper.

%%%%%%%%%%%%%%%%%%%%%%%%%%%%%%%%%%%%%%%%%%%%%%%%%%%%%%%%%%%%%%%%%%%%%%%%%%%%%%%%
\section{Results}\label{results}
\label{results_spac}
We have computed the probabilities $P^t(avg_i)$ and report those which feature below $5\%$ or above $95\%$, corresponding to departures from the $5\%$\textendash$95\%$ confidence range of hypothesis testing methodology \citep{doi:10.2307/3315916,fisher1925statistical,10.12688/f1000research.6963.5,Neyman:1937uhy}.
%\subsection{Spacings}\label{results_spac

\begin{figure*}%[0]{I}{\columnwidth}%[h!]
 \centering
 \includegraphics[width=1.5\columnwidth]{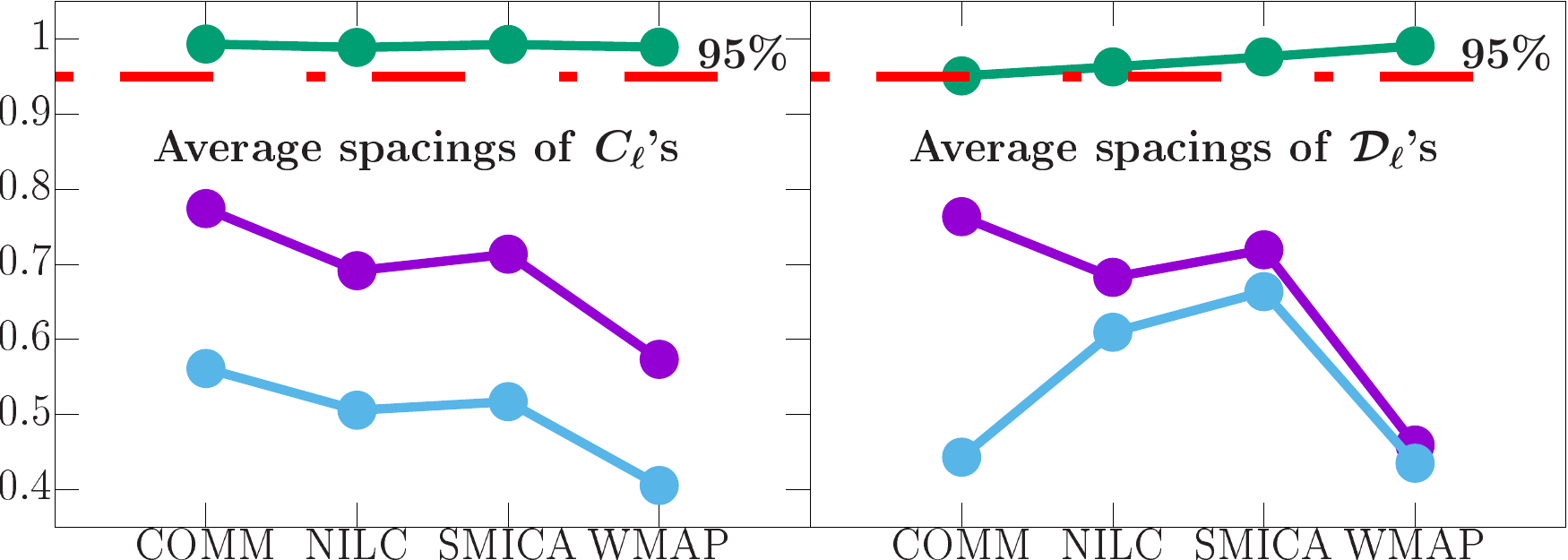}
\caption{Left panel: Probabilities $P^t(avg_a)$ (purple), $P^t(avg_e)$ (green), $P^t(avg_o)$ (light-blue) for $C_\ell$'s. Right panel: $P^t$ of average spacings of $\mathcal{D}_\ell$'s. Red dashed lines indicate $ 95\%$ C.L. The average spacings of even multipoles from cleaned CMB maps are unusually low.}
\label{fig_allfl1}
\end{figure*}

From the left panel of Figure \ref{fig_allfl1}, we see that the average spacings of $C_\ell$'s for all and odd multipoles are well within our confidence range. For even multipoles, however, all the four maps exhibit unusually low average spacings. From the right panel of Figure \ref{fig_allfl1}, again, we see that the average spacings of $\mathcal{D}_\ell$'s for all and odd multipoles are as expected, but those for even multipoles are unusually low for all the four maps. The values of $P^t$ for even multipoles are in Table \ref{tab1}.

\begin{table}
%\tiny
    \centering
    \caption{Values of $P^t(avg_e)$ for four maps, for $C_\ell$'s and $\mathcal{D}_\ell$'s.}
    \label{tab1}
    \vspace{0.4cm}
\begin{tabular}{|c|c|c|c|} 
%{\textbf{$f(\ell)=\frac{\ell(\ell+1)}{2\pi}$}} \\ [2ex]
\hline 
Map & for $C_\ell$'s & for $\mathcal{D}_\ell$'s \\ [1ex] 
 \hline%\hline
COMM & $99.33\%$ & $95.07\%$ \\ [1ex]
 %\hline
NILC &$98.86\%$& $96.27\%$  \\ [1ex]
 %\hline
SMICA & $99.26\%$ & $97.61\%$  \\ [1ex]
 %\hline
WMAP &  $98.90\%$ & $99.08\%$ \\ [1ex] 
 \hline
\end{tabular}
\end{table}

Overall, we see that the multipole spacings of even multipoles are unusually low relative to those based on the $\Lambda CDM $ concordance model, and rejected at $\gtrsim 95\%$ C.L. for all the four foreground cleaned CMB maps. This may mean that the unexpected signal is either truly characteristic of the CMB sky or that it has been left over due to a generic systematic error or a similar foreground residual in all the maps, despite the use of different cleaning methods. In either case, such occurrences must be further checked for their robustness. In the next section, we subject our findings to rigorous checks of robustness with the help of an inpainting method based on constrained Gaussian realisations with two kinds of galactic masks. In addition, we mask the non-Gaussian cold spot (NGCS) and inpaint over the same. Thus, evaluating the average spacing estimator for the foreground cleaned inpainted CMB maps helps us explore any variations of the signal with respect to (a) possibly minor foreground residuals in the galactic region, and (b) the NGCS.

\section{Robustness with Inpainted Realisations}\label{inpainted}

\begin{figure*}[htbp]
    \centering
    \includegraphics[width=1.5\columnwidth]{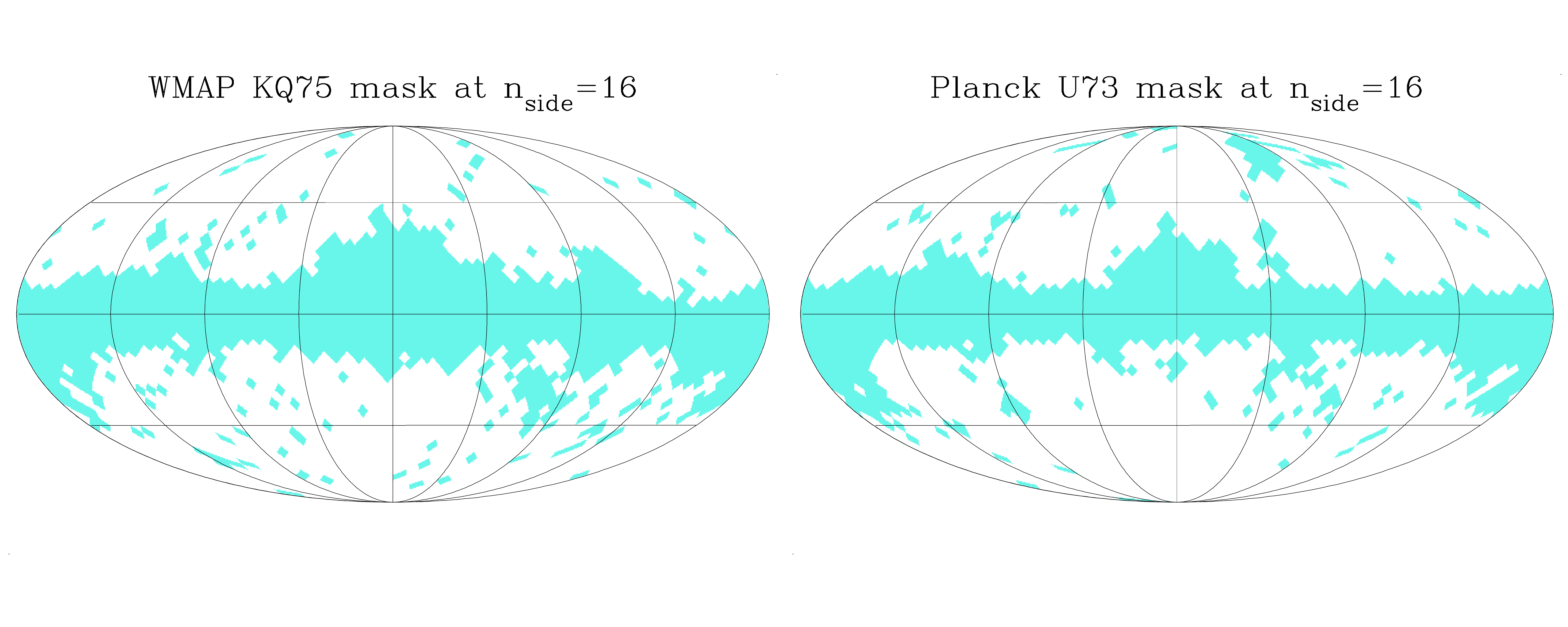}
    \caption{Low resolution (HEALPix $n_{side}=16$) versions of WMAP's $KQ75$ and Planck's $U73$ masks. Masked regions are indicated in cyan.}
    \label{masks}
\end{figure*}

The unusually low valued estimator ($avg_e$) found in the previous section albeit is seen consistently for all four maps, yet, the result could be due to foreground residuals or other unresolved systematics. To distinguish the occurrence of unusual patterns as being characteristic of the CMB sky as opposed to residual uncertainties, we use masks for the galactic region and some extragalactic point sources. Since foreground residuals in and around the galactic region may cause such unexpected signatures. The $KQ75$ mask of WMAP, and a product of the temperature confidence masks of COMM, NILC, SEVEM, and SMICA of Planck, referred to as the $U73$ mask, have been used here. These are shown in Figure \ref{masks}.

To obtain these low resolution (HEALPix $n_{side}=16$) masks, we downgraded the available high resolution masks and applied a threshold of $x=0.85$ to the $KQ75$ mask and that of $x=0.98$ to the $U73$ mask. This implies that after downgrading, we set all pixels with value $\leq x$ to $0$, and the others to $1$. Being a conservative mask, $KQ75$ includes a wider galactic cut and many point sources relative to the $U73$ mask. Choices for the thresholds are based on \cite{_iso_stat}, and help us keep a considerable sky fraction ($62.9\%$ for $KQ75$ and $67.5\%$ for $U73$) while masking regions with dominant foreground sources at large scales  \citep{Tegmark_2000}.  Hence the analysis for $avg_e$ was repeated on full sky inpainted realisations of the masked CMB maps.

\subsection{Inpainting method}
The inpainting method used is that of local constrained Gaussian realisations in pixel space in regions of the sky that are masked. This method is based on \cite{10.1111/j.1365-2966.2012.21138.x, refId0} and can be outlined in the following steps:

\begin{enumerate}
    \item Consider the entire set of pixels of an foreground cleaned CMB full sky map, as $m=\left(\begin{array}{c}
p_{obs}  \\
q_{obs}  \\
 \end{array}\right)$, where, $p_{obs}$ represents the set of pixels in a masked region, and $q_{obs}$ represents those of the unmasked region that will be used to constrain the pixels in the masked region.
 \item With a fiducial power spectrum $C_\ell^{fid}$, a full sky Gaussian random realisation is generated given by $m_r=\left(\begin{array}{c}
p_{r}  \\
q_{r}  \\
 \end{array}\right)$. Here $C_\ell^{fid}$ is the theoretical best fit to Planck 2018 data.
 \item Set $q_{in}=q_{obs}-q_{r}$
 \item Calculate covariance matrices $\mathcal{C}_{pq}, \mathcal{C}_{qq}$ of pairs of masked-unmasked  and unmasked-unmasked pixel sets, respectively, determined from random Gaussian realisations based on a theoretical ($\Lambda CDM$ model, say) power spectrum best fitted to some data (here, Planck 2018 data). A simpler method to calculate these covariance matrices, as used here is by considering the expression for elements of the CMB covariance matrix $\mathcal{C}$ \citep{sudevan_global_2018} :
 \begin{eqnarray}
 \mathcal{C}_{ij}&=&\sum\limits_{\ell=2}^{n\ell_{max}}\frac{2\ell+1}{4\pi}C_\ell^{fid} B_\ell^2P_\ell^2\mathcal{P}_\ell(\cos(\gamma_{ij})), \nonumber \\
 \cos(\gamma_{ij})&=&\cos(\theta_i)\cos(\theta_j)\nonumber\\
 &&+\sin(\theta_i)\sin(\theta_j)\cos(\phi_i-\phi_j),
 \end{eqnarray} where, $(\theta_i,\phi_i)$ are the spherical polar angles for the $i^{th}$ pixel, $B_\ell, P_\ell$ are the beam and pixel window functions, respectively, and $\mathcal{P}_\ell$ is the Legendre polynomial.
 \item Set $p_{ml}=\mathcal{C}_{pq}\mathcal{C}_{qq}^{-1} q_{in}$. The inverse can be found using the Moore-Penrose generalised pseudo inverse \citep{bams/1183425340, penrose_1955, sudevan_global_2018}. 
 \item Set $p_{cg}=p_{ml}+p_{r}$. Thus $p_{cg}$ corresponds to pixels of a local constrained Gaussian realisation.
\item The set of pixels for the complete inpainted map are given by $m_{cg}=\left(\begin{array}{c}
p_{cg}  \\
q_{obs}  \\
 \end{array}\right)$
\end{enumerate}

\begin{figure*}[tbp]
    \centering
    \includegraphics[width=1.5\columnwidth]{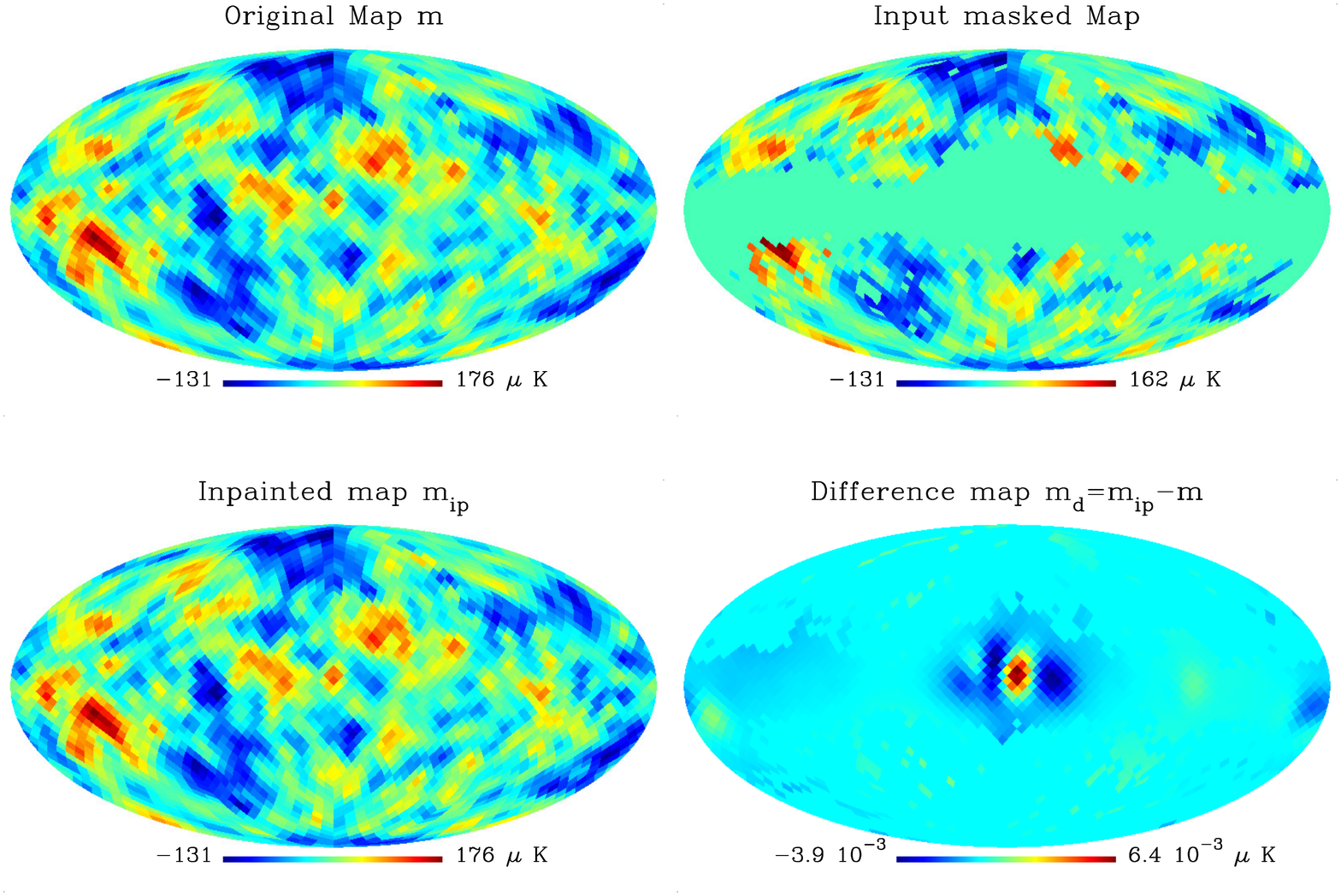}
    \caption{Top left panel: Original statistically isotropic CMB map $m$; top right panel: Map $m$ after masking; bottom left panel: map $m_{ip}$ after inpainting over the masked region; bottom right panel: difference between the original and inpainted maps. The difference map has values of order $10^{-3}$. This helps demonstrate the efficacy of the inpainting method used, which is that of a local constrained Gaussian realisation.}
    \label{demip}
\end{figure*}

To demonstrate the efficacy of the inpainting method used here, we have considered a statistically isotropic CMB realisation map and its inpainted version. The inpainting was done on the partial sky map obtained with a mask which is a product of the $KQ75$ and $U73$ masks, and hence a more conservative mask relative to either of the two. Beside the inpainted map, we have presented a Mollweide projection of their difference. Say, for the original map, given by $m$, and its inpainted version given by $m_{ip}$, the difference map is $m_d=m_{ip}-m$. So $m$, $m_{ip}$ and $m_{d}$, for a statistically isotropic CMB map are shown in Figure \ref{demip}.

\subsection{Results from inpainting over \textit{KQ75} and \textit{U73} masks}

After separately applying the $KQ75$ and $U73$ masks to cleaned CMB maps, we generated $10^3$ inpainted realisations of each of those, and have shown normalised counts ($\mathcal{N}$) of the $avg_e$ estimator. This quantity $\mathcal{N}(avg_e)$ is the number of realisations with a value of $avg_e$ in a certain bin, divided by the total number of realisations. Notably, these curves for both $C_\ell$'s and $\mathcal{D}_\ell$'s have very small spreads. Hence these resemble nearly vertical lines when shown along with $\mathcal{N}(avg_e)$ from statistically isotropic CMB realisations. We calculated $P^t(avg_e)$ for each of the $10^3$ inpainted realisations of a particular map and for a certain mask. We found that the values for all realisations of each of the four respective cleaned maps are the same as those in Table \ref{tab1}, regardless of the mask used.

\begin{figure*}[tbp]
    \centering
    %\vspace{cm}
    \includegraphics[height=1.15\textwidth]{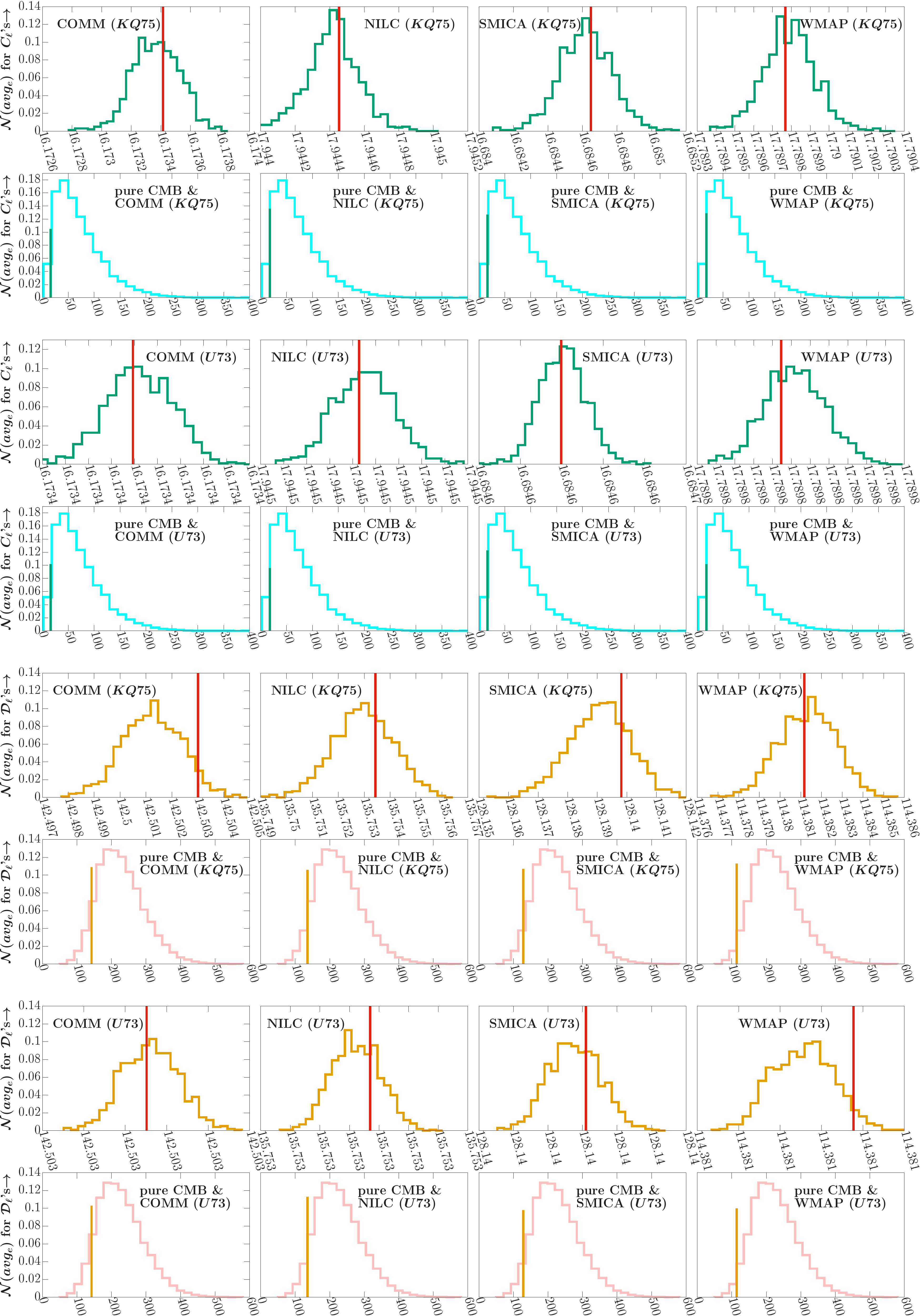}
    \caption{First row: Normalised counts $\mathcal{N}(avg_e)$ for $C_\ell$'s using $10^3$ inpainted realisations of COMM, NILC, SMICA, WMAP with the $KQ75$ mask. Second row: $\mathcal{N}(avg_e)$ as in first row, along with that from $10^4$ statistically isotropic CMB realisations. Third, fourth row: Same as first and second rows, but with the $U73$ mask.
    Fifth row: $\mathcal{N}(avg_e)$ for $\mathcal{D}_\ell$'s using $10^3$ inpainted realisations of COMM, NILC, SMICA, WMAP with the $KQ75$ mask. Sixth row: $\mathcal{N}(avg_e)$ as in fifth row, along with that from $10^4$ statistically isotropic CMB realisations. Seventh, eighth row: Same as first and second rows, but with the $U73$ mask.
     Vertical red lines indicate values from full sky cleaned maps.
    }
    \label{ip1}
\end{figure*}

In the first row of Figure \ref{ip1}, from left to right, the four subfigures correspond to $\mathcal{N}(avg_e)$ of $C_\ell$'s from inpainted realisations of COMM, NILC, SMICA and WMAP using the $KQ75$ mask (dark-green).  Vertical red lines indicate values of $avg_e$ from the four originally cleaned full sky maps. The second row shows the same curve of $\mathcal{N}(avg_e)$ (in dark-green) along with that from $10^4$ statistically isotropic CMB realisations (in cyan). The third and fourth rows show the same curves as the first and second rows, respectively but with the use of the $U73$ mask.

The fifth and sixth rows of Figure \ref{ip1} follow the same pattern as the first and second rows, but for $\mathcal{D}_\ell$'s. The curves of $\mathcal{N}(avg_e)$ for $\mathcal{D}_\ell$'s from $10^3$ inpainted realisations using the $KQ75$ mask are shown in orange, while that from $10^4$ statistically isotropic CMB maps is shown in pink. The seventh and eighth rows follow the same pattern as the fifth and sixth rows, but with the $U73$ mask.

From these subfgures we see that the spreads of $\mathcal{N}(avg_e)$ from inpainted realisations are very small and closely centred around the red lines from the full sky foreground cleaned CMB maps. Besides, some curves of $\mathcal{N}(avg_e)$ from inpainted realisations that are shifted significantly tend to favour lower values of $avg_e$ compared to these red lines (for e.g., fifth-row-first-column and seventh-row-fourth-column subfigures of Figure \ref{ip1}). Again relative to the curve from statistically isotropic CMB maps, these confirm that they lie on the leftmost unlikely regions, as seen before (Table \ref{tab1}). Thus inpainting over the two masks indicates that the signal of anomalously low $avg_e$ persists for both $C_\ell$'s and $\mathcal{D}_\ell$'s of $\ell\in[2,31]$, robustly for four different cleaning methods and two different masks. This makes it difficult to attribute the same to foreground residuals.

\subsection{Effect of the non-Gaussian cold spot (NGCS)}

\begin{figure}[tbp]
    \centering
    \includegraphics[width=0.75\columnwidth]{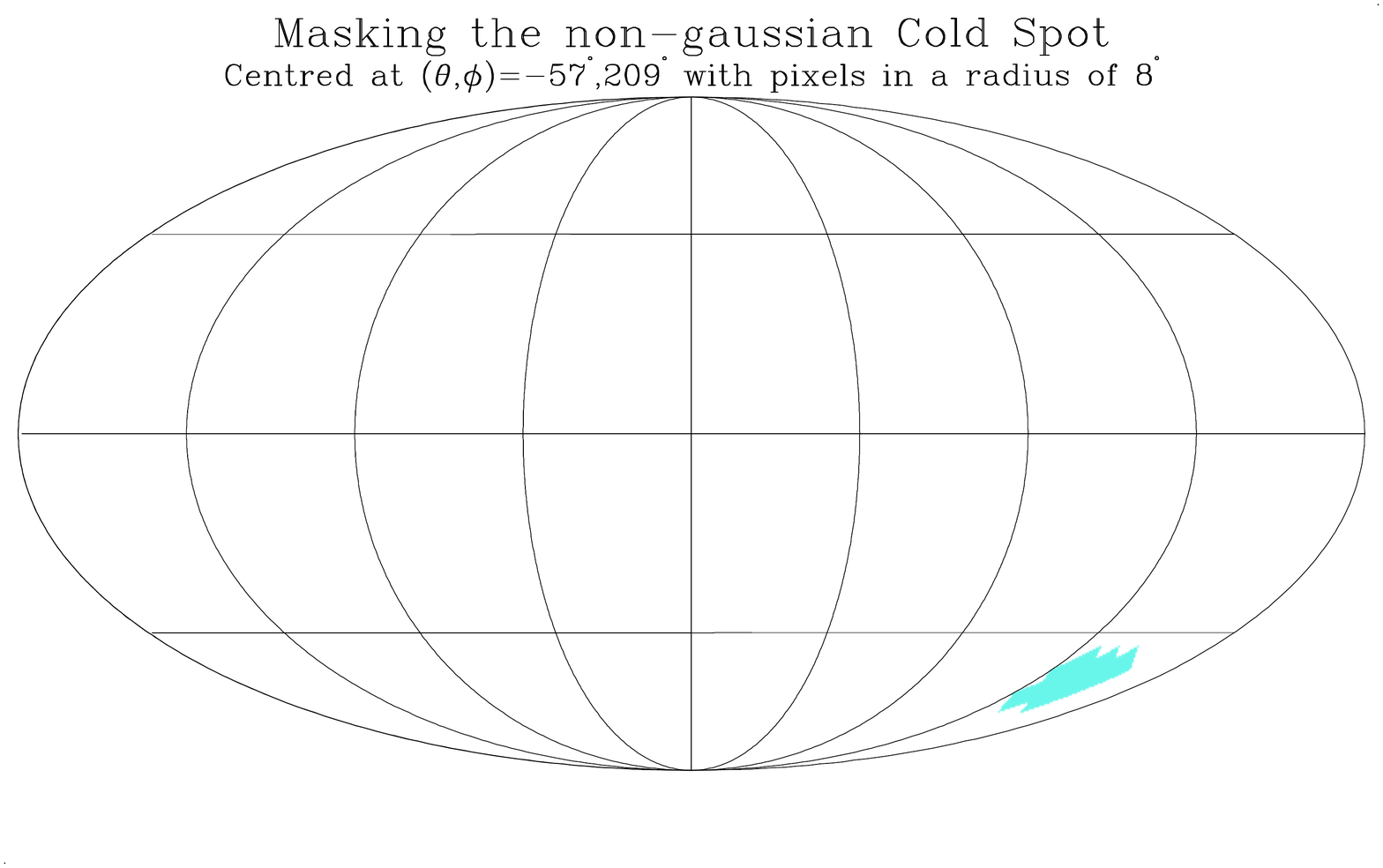}
    \caption{The non-Gaussian cold spot (NGCS) mask: the NGCS is shown in cyan; the unmasked region is in white.}
    \label{cs}
\end{figure}

The NGCS \citep{10.1111/j.1365-2966.2006.10312.x, MARTINEZGONZALEZ2006875, vielva_2010,PhysRevD.80.123010} approximately centred at $(\theta,\phi)=\ang{-57},\ang{209}$ was shown to be correlated with the north-south power asymmetry \citep{Bernui_2014, Quartin_2015, Akrami_2014, Eriksen_2004, Eriksen_2004b, Eriksen_2007}. However, the significance of this effect was seen to be low for low resolution \citep{10.1111/j.1365-2966.2010.16905.x} maps at HEALPix $n_{side}=16$. Nevertheless, a study of its relation with the deficit of large-angle correlations has not been investigated. We therefore generate a mask for the NGCS by setting all pixels in a radius of $\ang{8}$ from its center to zero, as shown in Figure \ref{cs}. We utilise this mask in union with the $KQ75$ and $U73$ masks (referred to as the $KQ75-CS$ and $U73-CS$ masks) and redo the analysis with such inpainted realisations of the four foreground cleaned CMB maps. 

\begin{figure*}[tbp]
    \centering
      %\vspace{-2cm}
    \includegraphics[height=1.15\textwidth]{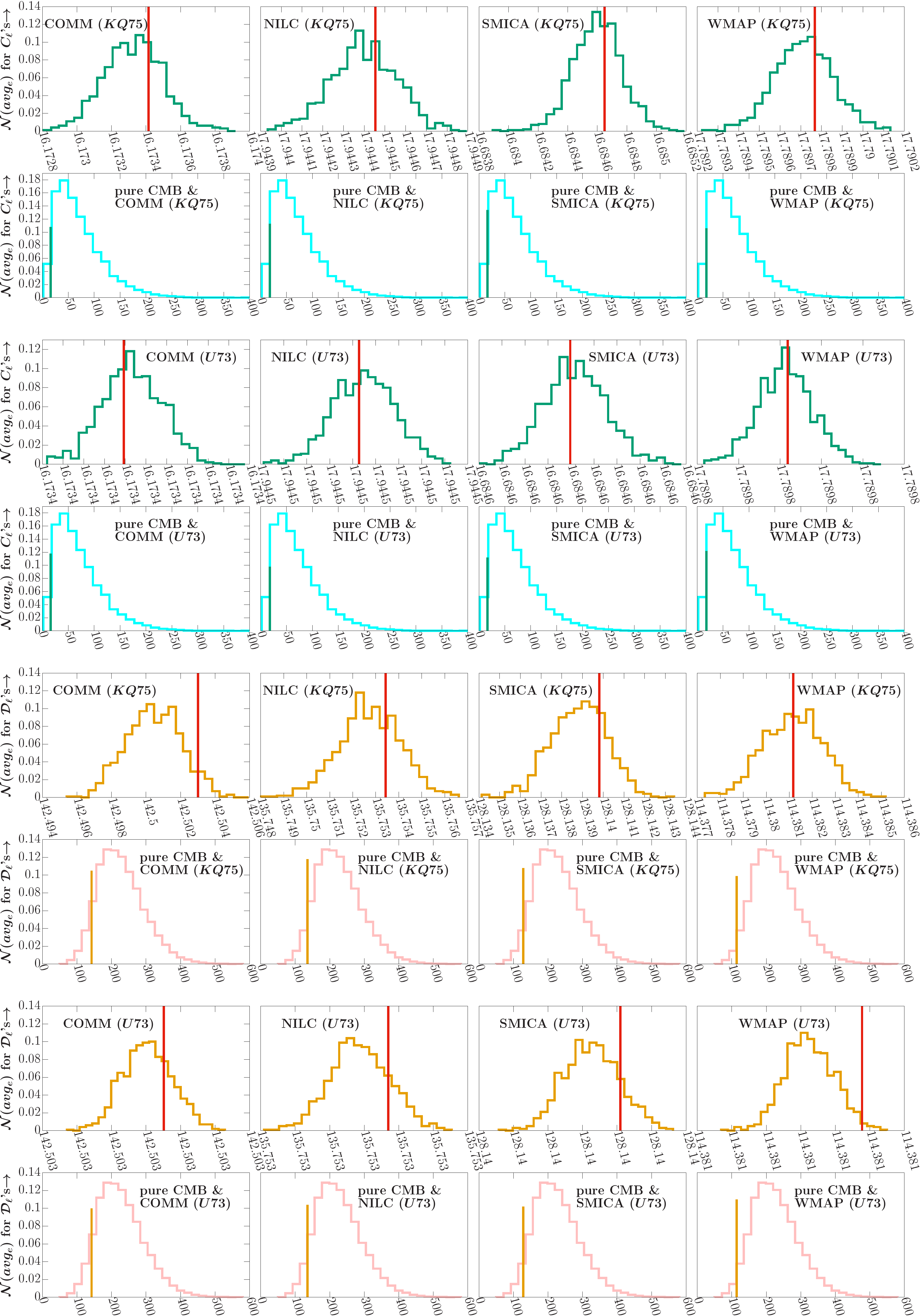}
    \caption{First row: Normalised counts $\mathcal{N}(avg_e)$ for $C_\ell$'s using $10^3$ inpainted realisations of COMM, NILC, SMICA, WMAP with the $KQ75-CS$ mask. Second row: $\mathcal{N}(avg_e)$ as in first row, along with that from $10^4$ statistically isotropic CMB realisations. Third, fourth row: Same as first and second rows, but with the $U73-CS$ mask.
    Fifth row: $\mathcal{N}(avg_e)$ for $\mathcal{D}_\ell$'s using $10^3$ inpainted realisations of COMM, NILC, SMICA, WMAP with the $KQ75-CS$ mask. Sixth row: $\mathcal{N}(avg_e)$ as in fifth row, with that from $10^4$ statistically isotropic CMB realisations. Seventh, eighth row: Same as first and second rows, but with the $U73-CS$ mask. Vertical red lines indicate values from full sky cleaned maps. We have omitted the suffix `$-CS$' due to lack of space in the subfigures.
    }
    \label{ip2}
\end{figure*}

In the first row of Figure \ref{ip2}, from left to right, the four subfigures correspond to $\mathcal{N}(avg_e)$ of $C_\ell$'s from inpainted realisations of COMM, NILC, SMICA and WMAP using the $KQ75-CS$ mask (in green). Vertical red lines indicate values of $avg_e$ from the original cleaned full sky maps. The second row shows the same $\mathcal{N}(avg_e)$ along with that from $10^4$ statistically isotropic CMB realisations (in cyan). The third and fourth rows show the same curves as the first and second rows, respectively but with the use of the $U73-CS$ mask. The fifth and sixth rows follow the same pattern as the first and second rows, but for $\mathcal{D}_\ell$'s. The seventh and eighth rows follow the same pattern as the third and fourth rows for $\mathcal{D}_\ell$'s.

The values of $P^t(avg_e)$ for these inpainted realisations are again the same as those of their respective full sky cleaned maps as in Table \ref{tab1}, regardless of the mask used. Thus, inpainting over the union masks, $KQ75-CS$ and $U73-CS$ indicates that the signal is independent of the NGCS. Thus a significantly low mean spacing of even multipole APS exists which is robust against four different cleaning methods, two different masks, and the presence or absence of the NGCS.

\section{Summary and Conclusion}\label{conclusion}

Level spacings of eigenvalues of random matrices have been studied before to classify the change of correlations between integrable and chaotic systems. Integrable systems are those for which the energy eigenvalues show level clustering (Poisson statistics) as they are uncorrelated, while those of chaotic systems show level repulsion (Wigner-Dyson statistics) due to presence of correlations.

Within the framework of the concordance model of cosmology, we expect the CMB to be statistically isotropic. This implies that the angular power spectrum (APS) of CMB is uncorrelated between different multipoles. Since foreground cleaned CMB maps are obtained from observed CMB radiation after application of various state of the art foreground cleaning methods, these maps are expected to be representative of the actual CMB sky, which is hypothesised to have no correlations in its APS measures. Thus it is interesting to probe the nature of any possible correlations in the APS of foreground cleaned maps to ascertain if the principle of statistical isotropy is obeyed. We note that a breakdown of statistical isotropy could be due to several possible mechanisms. These include the presence of a statistically anisotropic primordial signal, or some minor residual foregrounds, or any unaccounted agents between the source and the observer, or due to any minuscule systematics left over as a result of the analysis pipeline employed during satellite data collection and/or the map making procedure.

The presence or absence of correlations can be concretely established with the help of the mean gap ratio, which avoids the problem of unfolding. We show that in the context of simulated statistically isotropic CMB maps, the mean gap ratio closely corresponds to that for Poisson statistics, whereas, on introduction of statistical anisotropy in the maps, we see a shift towards some appropriate level repulsion statistics. The mean gap ratio is obtained by averaging over an ensemble of CMB realisations, similar to how quantities like the correlation coefficients are ascertained. Since we have only one CMB sky to observe instead of an ensemble, therefore we devise a novel estimator which computes the average APS spacing of a set of low multipoles ($\ell \in [2,31]$). We show that such an estimator can distinguish between statistically isotropic and anisotropic CMB, and hence is useful in categorising the nature of correlations present in foreground cleaned CMB maps. This estimator is computed for even and odd multipole spacings in addition to all multipoles taken together.

Without any parity distinction, for all multipoles, the spacings are seen to be in good agreement with theoretical expectation. Parity based distinction reveals that the average spacing of even multipoles ($avg_e$) is anomalously low for both $C_\ell$'s (at $\geq 98.86\%$ C.L.) and $\mathcal{D}_\ell$'s (at $\geq 95.07\%$ C.L.). Since all four maps, namely, COMM, NILC, SMICA, and WMAP, are obtained with the help of different foreground cleaning algorithms, the amounts of foreground residuals in these maps are different \citep{Larson_2015, article2, 2004ApJ...612..633E}. These systematic differences are distinctly visible if we consider any two of these maps at low resolution and subtract one from the other. Hence we perform further studies with inpainted realisations of masked CMB maps, to establish whether the observed anomalously low $avg_e$ spacings are due to foreground residuals, and we find that the signal persists in all the foreground cleaned inpainted CMB maps. We conclude that this signal of unusually low average even multipole spacings is robust against 
\begin{enumerate}
    \item[(a)] the use of two different galactic masks,
    \item[(b)] data from two different instruments, i.e, WMAP and Planck satellites,
    \item[(c)] consideration of maps obtained from four different cleaning methods, namely those of Gibbs sampling for COMM, Spectral Matching Independent Component Analysis for SMICA, Internal linear combination (ILC) in needlet space for NILC, ILC in pixel space for WMAP, and
    \item[(d)] the presence or absence of the non-Gaussian cold spot.
\end{enumerate}

Thus, we find a robust signal of low average spacing for $\ell \in [2,31]$ with even multipoles of $C_\ell$'s and $\mathcal{D}_\ell$'s which seems unlikely due to foreground residuals in the galactic region of cleaned maps. This finding is in agreement with previous findings of the deficit of large-angle correlation and its equivalence with the odd-parity preference of the APS. However, our findings may additionally indicate that correlations between odd multipole APS are not anomalous, as opposed to those of even multipole APS. This accounts for a possibly unusual level clustering of even multipole APS or a spacing distribution that favours low even multipole spacings. The unusually low average even multipole spacing hints at possible breakdown of statistical isotropy which is primordial in origin. For instance, there could be the possibility of an anisotropic Finsler spacetime model \citep{10.1093/mnras/sty1689} with a correction term that lowers even multipole $C_\ell$'s. A theoretical model that alters the correlations of primordial fluctuation modes \citep{PhysRevD.79.043015}, could inspire some alternate work to shed light on the low even multipole APS correlation.

\section*{Acknowledgements}

We acknowledge the use of the publicly available HEALPix \citep{2005ApJ...622..759G} software package (\url{http://healpix.sourceforge.io}). Our analyses are based on observations from Planck (\url{http://www.esa.int/Planck}), an ESA science mission with instruments and contributions directly funded by ESA Member States, NASA, and Canada. We acknowledge the use of the Legacy Archive for Microwave Background Data Analysis (LAMBDA), part of the High Energy Astrophysics Science Archive Center (HEASARC). HEASARC/ LAMBDA is a service of the Astrophysics Science Division at the NASA Goddard Space Flight Center. MIK would like to thank Ujjal Purkayastha for help with the basics of HEALPix relevant to this project. MIK is grateful to Aamna Ahmed for discussions regarding mean gap ratios.

%%use \balance somewhere in the left column of the last page to balance the two columns in the end page

%%References section
\bibliography{references}
\bibliographystyle{apj}

\end{document}